\documentclass[prb,twocolumn,showpacs,amsmath,amssymb,superscriptaddress,floatfix]{revtex4}
\newcommand{\bra}[1]{\langle #1|}
\newcommand{\ket}[1]{|#1\rangle}

\usepackage{graphicx}
\usepackage{bm}
\usepackage[dvips]{color}

\begin{document}

\title{Switchable resonant coupling of flux qubits}
\author{M.~Grajcar}
\affiliation{Frontier Research System, The Institute of Physical
and Chemical Research (RIKEN), Wako-shi 351-0198, Japan}
\affiliation {Department of Solid State Physics, Comenius
University, SK-84248 Bratislava, Slovakia.} \affiliation {Center
of Excellence of the Slovak Academy of Sciences (CENG), Slovakia}
\author{{Yu-xi} Liu}
\affiliation{Frontier Research System, The Institute of Physical
and Chemical Research (RIKEN), Wako-shi 351-0198, Japan}
\author{Franco Nori}
\affiliation{Frontier Research System, The Institute of Physical
and Chemical Research (RIKEN), Wako-shi 351-0198, Japan}
\affiliation{MCTP, Physics Department, CSCS, The University of
Michigan, Ann Arbor, Michigan 48109-1040, USA}
\author{A.M. Zagoskin}
\affiliation{Frontier Research System, The Institute of Physical
and Chemical Research (RIKEN), Wako-shi 351-0198, Japan}
\affiliation{Physics and Astronomy Dept., The University of
British Columbia,    Vancouver, V6T 1Z1
Canada}
\date{\today}

\begin{abstract}
We propose a coupling scheme, where two or more flux qubits with different
eigenfrequencies share Josephson junctions with a coupler loop devoid of its
own quantum dynamics. Switchable two-qubit coupling is realized by tuning the
frequency of the AC magnetic flux through the coupler to a combination
frequency of two of the qubits. The coupling allows any or all of the qubits to
be simultaneously at the degeneracy point and can change sign.

\end{abstract}


\maketitle

\section{Introduction} The  scheme by Makhlin et al.\cite{Makhlin99} for
coupling superconducting qubits was recently followed by several  proposals for
{\em tunable }
coupling\cite{You02,You03,Filippov03,Averin03,Kim04,Plourde04,Castellano05,Brink05}
between superconducting qubits.\cite{You05} Broadly speaking, these approaches
couple qubits through the exchange of {\em virtual} excitations in the coupler
circuit, in which the energy separation $E_0$ between the ground and first
excited state    is much larger than the tunnel  splitting $\Delta$ in the
qubits. The coupling strength $J$ is controlled by tuning the energy of the
coupling circuit, via a magnetic field (if the coupler is a SQUID), or via a
gate voltage (if it is a Cooper pair box). 

In principle, the above approaches provide a DC coupling between qubits, and
they allow the realization of entangling gates only if the difference
$|\Delta_a-\Delta_b|$ between the tunnel splittings of the corresponding qubits
is smaller than the coupling energy $J$.

In the opposite limit, when  $|\Delta_a-\Delta_b|\gg |J|$, it was
shown\cite{Liu06} that  the qubit-qubit interaction can be controlled by an
external variable-frequency magnetic field at the combination frequencies,
$|\Delta_a \pm \Delta_b|/\hbar$.   This approach\cite{Liu06} (coupling by using
a time-dependent magnetic flux, or TDMF)  is  advantageous due to the resonant
character of the coupling: in experiments it is often easier to produce fast and
precise frequency shifts    of the RF control signal,  as opposed to changes in
the amplitude of the DC signal. The proposal in Ref.~\onlinecite{Liu06} also did
not require additional, dedicated coupler circuits.  Its disadvantage was that
at least one of the qubits must be biased away from the optimal point, which
could make its operation more difficult and reduce its decoherence time.

Later on, a combination of the TDMF approach with a dedicated coupler circuit
led    to tunable-coupling  proposals in Refs.~\onlinecite{Bertet06} and
\onlinecite{Niskanen06}, where both coupled qubits could be simultaneously at
their optimal points.\footnote{However, Ref.~\onlinecite{Niskanen06}
results use the transformation $H' = UHU^{\dagger}$ without an additional term,
that is required when $U$ is time-dependent (which is the case).}  

In this paper we propose an alternative realization  of the TDMF coupling,
which allows  to {\em switch the coupling on and off, and to change its sign}.
Our proposal has an advantage over both the  approaches of
Refs.~\onlinecite{Liu06,Bertet06,Niskanen06}, and the generalization of
Ref.~\onlinecite{Niskanen06} to the Josephson coupling (described in Section
\ref{sec_sui_generis} of this paper), in {\em simultaneously} providing:  (1) a
coupling for arbitrarily biased qubits, (2) a higher coupling energy, (3)
enhanced protection from the flux noise, and (4) the elimination of  the
parasitic first-order DC coupling. 

\begin{figure}[h!]
\ \\ \  
\centerline{\includegraphics[width=7cm]{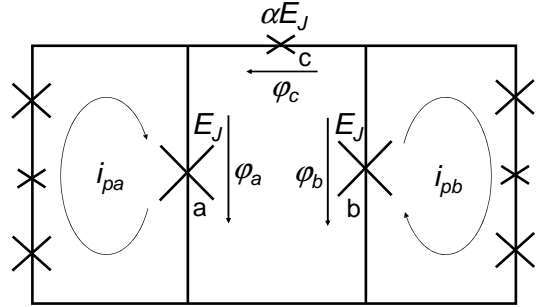}}
\caption{Schematic diagram of two flux qubits and the three-junction 
coupler circuit.}
\label{Fig:schem}
\end{figure}
\section{Model} In our proposal, the coupling circuit is a small-inductance
superconducting loop ($L_c\rightarrow 0$) with three Josephson junctions
(denoted by {\em a, b, c} in Fig.~\ref{Fig:schem}). The shared junctions {\em a,
b} ensure a significantly stronger  qubit-loop coupling than in the case of
purely inductive (like in Refs.~\onlinecite{Bertet06,Niskanen06}) or galvanic
connection.\cite{Plourde04} A
controllable DC coupling in a similar device  has been recently proposed and
realized experimentally,\cite{Kim04,Brink06,Ploeg06} with the coupling energy
$J_{\rm DC} = 1.7$~GHz. 

For the sake of simplicity, and without loss of generality,  we assume that the
junctions $a$ and $b$ have the same Josephson energy $E_J=\Phi_0I_c/2\pi$. The
qubit-qubit coupling is realized by the (small) junction $c$, with the Josephson
energy $\alpha E_J \ll E_J$. The coupler circuit ($a$, $b$, and $c$) has a
high plasma frequency, $\omega_p\sim\sqrt{8E_JE_C}/\hbar$, so that its
energy-level separation is much larger than all relevant characteristic energies
($J$ and $\Delta_{a,b}$) of the system. The large Josephson energy and large
capacitances ($C\gg C_c \approx \alpha C$) of the coupling junctions ensure that
$E_C/E_J\ll 1$. This allows us to neglect their degrees of freedom and to
consider them as passive elements, which convert the bias currents $I_{pa}$ and
$I_{pb}$, produced by the persistent currents circulating in the attached flux
qubits, into the phase  shift, and therefore the energy shift, of the
small Josephson junction $c$.

\section{Coupling strength} 
Let us now concentrate on the coupler circuit. The action of the two flux qubits
on it can be represented by the bias currents through the junctions $a$ and $b$.
(Due to the dominance of the Josephson coupling, we can disregard the geometric
mutual inductances between them.) The qubit-qubit interaction energy is obtained
by taking into account the total potential energy. The latter is the free
energy of the coupler plus the work performed by the qubits $a,b$ on the coupler
circuit to keep their persistent currents constant.\cite{Likharev} By making use
of the quantization condition for the gauge-invariant phase differences,
$\varphi_a-\varphi_b+\varphi_c=-2\pi \Phi_c/\Phi_0$, where $\Phi_c$ is the
magnetic flux through the coupler loop and $\Phi_0$ is the magnetic flux
quantum, the reduced total energy $U_t$ can be written  as
\begin{eqnarray}
\tilde{U}_t  \:\equiv\: U_t/E_J \:= -\cos\varphi_a-\cos\varphi_b\nonumber \\
-\alpha\cos(2\pi
f_c+\varphi_a-\varphi_b)-i_{pa}\,\varphi_a + i_{pb}\,\varphi_b,
\end{eqnarray}
where  $f_c=\Phi_c/\Phi_0$, $i_{pa}=I_{pa}/I_c$, and
$i_{pb}=I_{pb}/I_c$. For small values of $\alpha,$ $i_{pa},$ and $
i_{pb}$, this potential forms a well with a minimum near the point
$(\varphi_a,\varphi_b)=(0,0)$. This is the Hamiltonian of a three-junction flux
qubit\cite{Orlando99} with biased junctions, which can be reduced to the
Hamiltonian   of a perturbed two-dimensional oscillator  
\begin{eqnarray}
H&=&\frac{\hat{P}_+^2}{2M_+}+\frac{\hat{P}_-^2}{2M_-}
+E_J\varphi_+^2+\kappa E_J(\varphi_--\varphi_-^*)^2
\nonumber\\
&+&E_J[\varphi_+(i_{pb}-i_{pa})-\varphi_-(i_{pa}+i_{pb})]. \label{eq_ref_1}
\end{eqnarray}
Here $\varphi_\pm=(\varphi_a\pm\varphi_b)/2$, 
$\hat{P}_\pm=i\hbar\partial/\partial\varphi_\pm$,  $M_+=2C(\Phi_0/2\pi)^2$,
$M_-=M_+(1+2\alpha)$, $\kappa=1+2\alpha\cos(2\pi f_c+2\varphi_-^*)$,  and 
$\varphi_-^*=-\alpha\sin(2\pi f_c)/(1+2\alpha\cos(2\pi f_c))$.
(see Ref.~\onlinecite{Orlando99} for details). The perturbation is
$E_J[\varphi_+(i_{pb}-i_{pa})-\varphi_-(i_{pa}+i_{pb})]$.

From Eq.~(\ref{eq_ref_1}), the normal frequencies of the coupler are
$\omega_+=\sqrt{2E_J/M_+}$ and $\omega_-=\sqrt{2\kappa E_J/M_-}$. Its
eigenstates, in the lowest order in $i_{pa},$ $i_{pb}$, are products
$\Psi_{+,m}(\varphi_+)\Psi_{-,n}(\varphi_-)$ of the eigenstates of the normal
modes. The first-order correction to the ground state energy of the coupler is
zero, and the coupling energy is determined by the second-order
correction:
\begin{eqnarray}
E_0^{(2)}&=&\frac{E_J^2(i_{pa}+i_{pb})^2}{\hbar\omega_-}
\left|
\bra{\Psi_{-,1}}\varphi_-\ket{\Psi_{-,0}}\right|^2\nonumber \\
&+& \frac{E_J^2(i_{pa}-i_{pb})^2}{\hbar\omega_+}
\left|\bra{\Psi_{+,1}}\varphi_+\ket{\Psi_{+,0}}\right|^2. \label{eq_ref_3}
\end{eqnarray}
Thus, it is evident that the coupling  is provided by the virtual photon
exchange between the qubits and the coupler. 

Separating the term proportional to
$i_{pa}i_{pb}$ in the second-order correction in Eq.~(\ref{eq_ref_3}), 
we obtain the coupling energy 
\begin{equation}
\label{Eq:Jk}
J=\frac{\kappa-1}{\kappa}E_J\frac{i_{pa}i_{pb}}{2}.
\end{equation}
Inserting the definition of $\kappa$ into Eq.~\ref{Eq:Jk} the
coupling energy of the three-junction coupler reads
\begin{equation}
J=\frac{\alpha
E_J\;\cos(2\pi f_c\; +\; 2\varphi_-^*)}{1+2\alpha\cos(2\pi f_c\; +\; 2\varphi_-^*)}\;i_{pa}i_{pb}.
\label{Eq:J}
\end{equation}
This expression  obviously corresponds to the $\sigma_z\sigma_z$ coupling in
the natural basis of qubit states (see, e.g.,  Eqs.~(1,3) in Ref.~\onlinecite{Ploeg06})
\begin{equation}
H_{\rm int}(t) = J(f_c)\,\sigma^z_a\,\sigma^z_b. \label{eq_ref_2}
\end{equation}
Obviously, the coupling (\ref{eq_ref_2}) allows either one or both qubits to be
in   their optimal points. In the experiment in Ref.~\onlinecite{Ploeg06}
such interaction was used to realize a tunable DC coupling between qubits $a$
and $b$, by changing the coupler bias $f_c$. The strength and sign of the
coupling depend on the precise value of $f_c$. As  mentioned above, for
time-domain operations it is often easier to manipulate the {\it frequency} of
the AC signal $f_c(t)$, rather than the {\it amplitude} of  a DC pulse.  We will
therefore use the TDMF approach initially proposed in Ref.\onlinecite{Liu06}.

\section{Effective coupling under TDMF}

Let us first consider the effective coupling for an arbitrary
$J(f_c)$.  Assuming the harmonic flux dependence, $$f_c(t)  = \nu_0
+ \nu_1 \cos\Omega t,$$ for the reduced flux in the coupler circuit, and expanding $J(f_c)$ near $\nu_0$, we 
reduce the Hamiltonian of the system to

\begin{eqnarray}
H(t) &=& H_0 + H_1 = \nonumber \\
&-&\frac{1}{2}\sum_{s=a,b} \Delta_s\sigma^x_s + \left(J_{\rm DC}(\nu_0)
+ J'(\nu_0) \nu_1\cos\Omega t \right)\sigma^z_a\sigma^z_b,\nonumber \\
\end{eqnarray}
where $J'(\nu_0)$ is the first derivative of the coupling energy, taken
at $\nu_0$. In the interaction representation this becomes
\begin{equation}
\tilde{H}(t) = H_0 + \tilde{H}_1(t)
\end{equation}
with
\begin{widetext}
\begin{equation}
\tilde{H}_1 = \left[J_{\rm DC}(\nu_0) + (J'(\nu_0) \nu_1\cos\Omega t)\right]
 (\sigma^z_a \cos\Delta_at - \sigma^y_a \sin\Delta_at)
(\sigma^z_b \cos\Delta_bt - \sigma^y_b \sin\Delta_bt). \label{eq_ref_sqbq}
\end{equation}
\end{widetext}

Assuming $\Delta_a-\Delta_b >0$ and $\Omega \approx
\Delta_a\mp\Delta_b$, we see that (after averaging over the fast
oscillations) only the coupling
\begin{equation}
H_{\rm eff} = \frac{J_{\rm AC}}{4}
\left[\sigma^z_a\sigma^z_b\pm\sigma^y_a\sigma^y_b\right]
\end{equation}
survives, where 
\begin{equation} J_{\rm AC}=J'(\nu_0) \nu_1. \label{eq_ref_X}
\end{equation}
The operator in the square brackets,  
 \begin{eqnarray}
 \sigma^z_a\,\sigma^z_b\pm\sigma^y_a\,\sigma^y_b \: = \:
\left(\begin{array}{rrrr} 1 & 0 & 0 & -1 \\
0 & -1 & \pm1 & 0\\
0 &  \pm1 & -1 & 0\\
-1 & 0 & 0 & 1 \end{array} \right),
\end{eqnarray}
 is entangling and therefore can be used to construct universal quantum ciruits.\cite{Zhang2003}

\begin{figure}
\centering
\includegraphics[width=8cm]{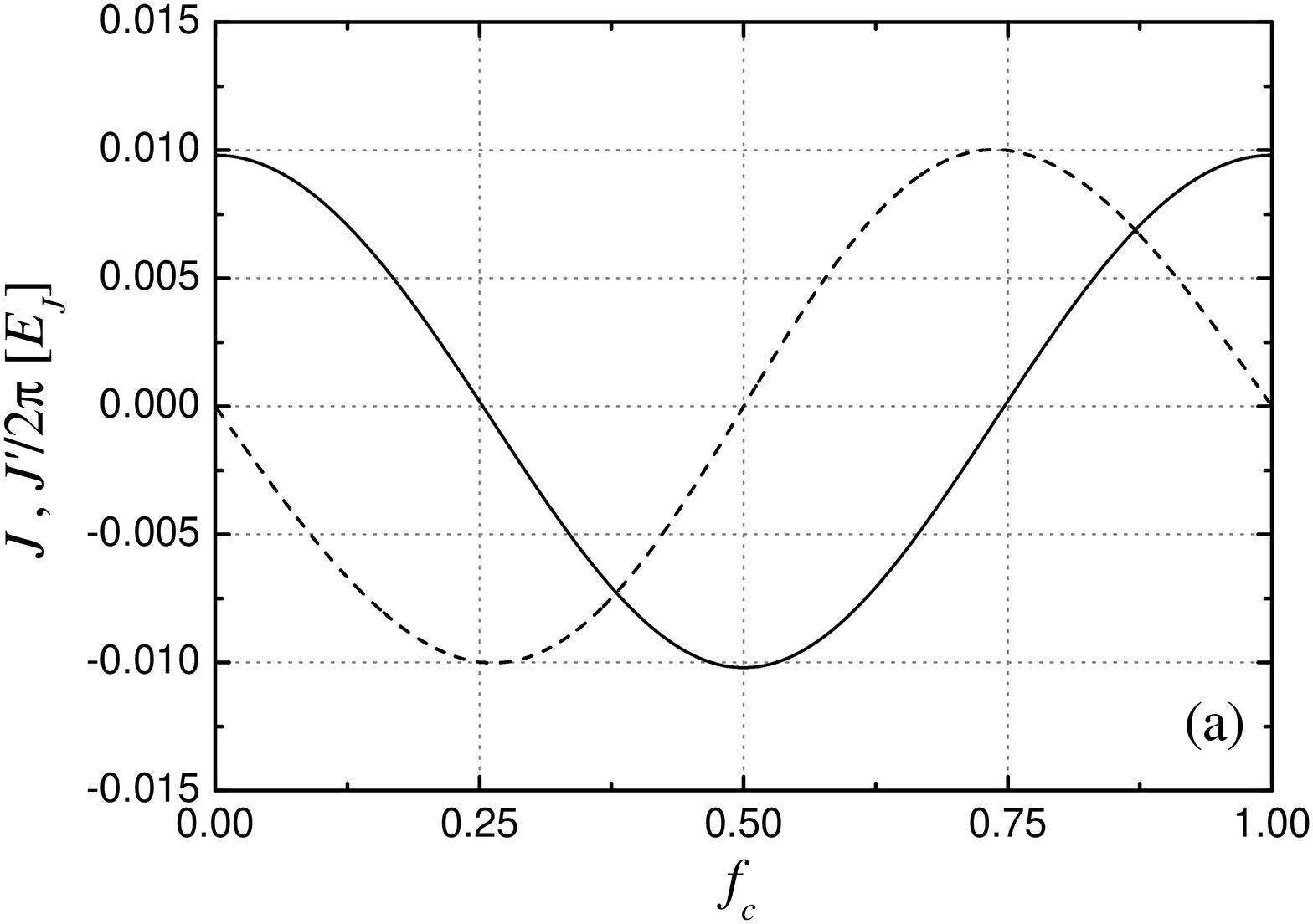}
\includegraphics[width=8cm]{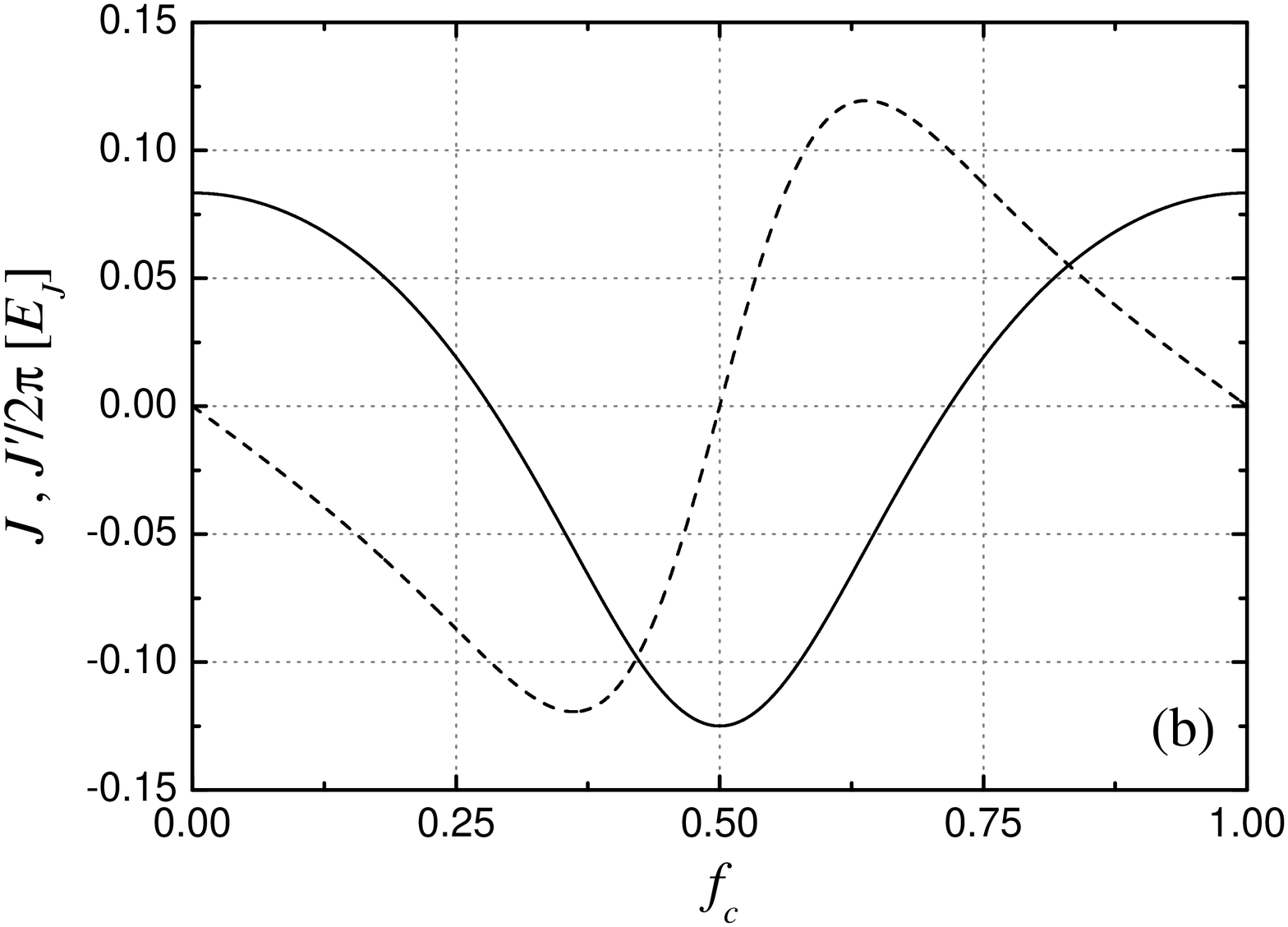}
\caption{Coupling energy $J$ (solid line) and its first derivative $J'$ (dashed line)
as a function of the reduced applied magnetic flux $f_c$ through the coupler loop,
for two parameters $\alpha=0.01$ (a) and $\alpha=0.1$ (b). 
The $J'$ is scaled by a factor $2\pi$. When the nonlinear response
of the coupler increases, the points of ``zero'' DC coupling and maximal AC coupling
diverge from the common point $f_c=0.25$.}
\label{Fig:JJ}
\end{figure}

Our results are somewhat similar to those of Ref.~\onlinecite{Niskanen06}. Let
us describe the differences. Due to the Josephson, rather than inductive,
coupling, our approach  realizes larger coupling energies, therefore it allows
smaller values of the parameter $\alpha$  and, correspondingly, is less
nonlinear. For example, at $\alpha=0.01$ and  $\nu_0=0.25$ the DC coupling
$J_{\rm DC}(0.25)$ is close to zero and the AC coupling (Eq.~(\ref{eq_ref_X}))
$J'(0.25)=2\pi E_{jc}i_ai_b$ is at a maximum (Fig.~\ref{Fig:JJ}a). Using the
experimental value of the  DC coupling energy for the device shown in
Fig.~\ref{Fig:schem}, $J_0(0)=1.7$~GHz,\cite{Ploeg06} we find the AC coupling
energy $J_{\rm AC}=10^{-2}J'(0.25)\approx 100$~MHz (for the reduced magnetic
flux amplitude $\nu_1=10^{-2}$).   The DC and AC couplings can be increased by
moving to the highly nonlinear regime with larger $\alpha >10^{-1}$, but now
the points, corresponding to ``zero'' DC coupling and maximal AC coupling, do
not coincide (Fig.~\ref{Fig:JJ}b).  

\section{Time-dependent magnetic flux coupling using an itermediary ``qubit"} \label{sec_sui_generis}

\begin{figure}
\centering
\includegraphics[width=7cm]{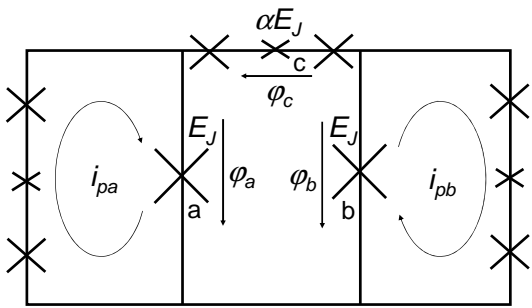}
\caption{Schematic diagram of two flux qubits and the five-junction (``qubit") coupler circuit.}
\label{Fig:schem2}
\end{figure}

Now let us substitute the smaller junction by
three Josephson junctions with sizes much smaller than the size of the coupling
junctions $a, b$ (see Fig.~\ref{Fig:schem2}). This is a generalized model of
Ref.~\onlinecite{Niskanen06}, 
with the inductive coupling replaced by the stronger
Josephson one. This  allows to increase   $\Delta_c$ and decrease the area of the qubits,
improving the  protection of the system against magnetic flux noise. 
Nevertheless we will see that this scheme is at a serious
disadvantage compared to the coupling of Fig.~\ref{Fig:schem}, because it leads
to a strong DC coupling between the qubits (i.e., parasitic DC coupling).

We can now apply the same approach  as in Eq.(\ref{eq_ref_3}). 
The  harmonic approximation of (\ref{eq_ref_1}) is now invalid. Instead,
the coupling energy is  determined by the change of the ground state energy of
the coupling ``qubit'' $c$,
$$\varepsilon(\varphi_c)=-\frac{1}{2}\sqrt{[2E_{Jc}(\varphi_c-\pi)]^2+\Delta_c^2}.$$
Here $E_{Jc}=\Phi_0
I_{pc}/2\pi$ and $\varphi_c$ is the phase difference across the
``qubit" $c$.   Expanding $\varepsilon(2\pi
f_c+2\varphi_-)$ at $2\pi f_c$ to  second order, we obtain the
potential of a two-dimensional linear harmonic
oscillator with a new value of the constant
$$\kappa=1+2\left(\frac{\partial^2\tilde{\varepsilon}}{\partial\varphi^2_c}\right),$$ where 
$\tilde{\varepsilon}=\varepsilon/E_J$ is the normalized ``qubit'' energy.
Substituting the new $\kappa$ in Eq.~(\ref{Eq:Jk}), we arrive at an expression 
for the coupling energy in the simple form 
\begin{eqnarray}
J&=&\frac{\partial^2\varepsilon}{\partial\varphi_c^2}\;i_{pa}i_{pb}\nonumber \\
&=&\frac{\Delta_c}{8\pi^2}\left(\frac{2I_{pc}\Phi_0}{\Delta_c}\right)^2
\left(1+\left(\frac{2I_{pc}\Phi_0}{\Delta_c}\tilde{f_c}\right)^2\right)^{-3/2}i_{qa}i_{qb},\nonumber \\
\end{eqnarray}
where $\tilde{f_c}=f_c-0.5$. The derivative, $J'=\partial J/\partial \tilde{f_c}$ has a maximum at
$\tilde{f}_{\rm cm}=\Delta_c/4I_{pc}\Phi_0$:  
\begin{equation}
J'_{\rm max}\approx J_{\rm DC}(f_{cm})\:\left(\frac{2I_{pc}\Phi_0}{\Delta_c}\right). 
\label{Eq:J_max}
\end{equation}
Near this point, the AC coupling energy depends on the external magnetic flux
only in the second order. This formula is equivalent to the expression (25) in
Ref.~\onlinecite{Niskanen06} provided that we use the standard normalization for
currents\cite{Likharev} $i_{pa}=2\pi M_{ac}I_{pa}/\Phi_0$, $i_{pb}=2\pi
M_{bc}I_{pb}/\Phi_0$ and neglect the mutual inductance between qubits.  It is
evident from Eq.~(\ref{Eq:J_max}) that the DC coupling {\em cannot be tuned to
zero} without switching off the AC coupling, and it turns out  to be much
stronger than the latter. This is what we refer to as parasitic DC coupling.
Because of large nonlinearity, the AC magnetic flux should be much smaller than
$\Delta_c/2I_{pc}\Phi_0$. Taking, e.g., $\nu_1=10^{-2}\Delta_c/2I_{pc}\Phi_0$,
the AC coupling energy becomes
\begin{equation}
J_{\rm AC} = J'_{\rm max}\;\nu_1 = 10^{-2}J(f_{\rm cm}),
\end{equation}
i.e., $J_{\rm AC} = 10^{-2}J_{\rm DC}.$ More importantly, in our proposal  {\em the DC
coupling can be switched off completely},  and the AC coupling (100 MHz) is
{\em five times stronger} than in Ref.~\onlinecite{Niskanen06}. 

\section{Conclusions}

We propose a feasible switchable coupling between superconducting flux qubits,
controlled by the resonant RF signal. Due to the frequency control, it is
particularly suitable for time-domain operations with flux qubits. The coupling
energy 100~MHz can be achieved by applying  a magnetic flux $10^{-2}\Phi_0$ to
the coupler with the combination frequency $$\omega_0 =
|\Delta_a\pm\Delta_b|/\hbar.$$ The Josephson coupling allows to minimize the
area of the devices, thus limiting the effects of the flux noise, and the
coupler thus can act in an almost linear regime, which, in particular,
suppresses the parasitic DC coupling. The resulting interaction term also acts
as an entangling gate and enables the realization of a universal quantum
circuit. 

\acknowledgments Authors thank Y. Nakamura and A. Maassen van den Brink for valuable discussion.
This work was supported in part by the Army Research
Office (ARO), Laboratory of Physical Sciences (LPS) and
National Security Agency (NSA); and also supported by the National Science Foundation
grant No. EIA-0130383. A.Z. acknowledges partial
support by the Natural Sciences and Engineering Research
Council of Canada (NSERC) Discovery Grants
Program and M.G. was partially supported by  Grants VEGA 1/2011/05 and APVT-51-016604.
\bibliography{ref_qubit}

\end{document}